# State-Space Inference for Non-Linear Latent Force Models with Application to Satellite Orbit Prediction

**Jouni Hartikainen**  JOUNI.HARTIKAINEN@AALTO.FI
Aalto University, Dept. of Biomedical Engineering and Computational Science, Espoo, Finland

**Mari Seppänen**  MARI.J.SEPPANEN@TUT.FI
Tampere University of Technology, Dept. of Mathematics, Tampere, Finland

**Simo Särkkä**  SIMO.SARKKA@AALTO.FI
Aalto University, Dept. of Biomedical Engineering and Computational Science, Espoo, Finland

## Abstract

Latent force models (LFMs) are flexible models that combine mechanistic modelling principles (i.e., physical models) with non-parametric data-driven components. Several key applications of LFMs need non-linearities, which results in analytically intractable inference. In this work we show how non-linear LFMs can be represented as non-linear white noise driven state-space models and present an efficient non-linear Kalman filtering and smoothing based method for approximate state and parameter inference. We illustrate the performance of the proposed methodology via two simulated examples, and apply it to a real-world problem of long-term prediction of GPS satellite orbits.

## 1. Introduction

Gaussian processes (GPs) are stochastic processes, which are commonly used for representing uncertainties of dynamic systems in many applications such as tracking, navigation and automatic control systems (Jazwinski, 1970; Bar-Shalom et al., 2001; Grewal & Andrews, 2001; Maybeck, 1982). In these applications, the Gaussian processes are typically white, and the processes are used as stochastic inputs in physical models such as driving forces of mechanical systems described in terms of ordinary differential equations (ODEs).

In a machine learning context, Gaussian processes (Rasmussen & Williams, 2006) are used as non-parametric models for unknown functions. Prior information on the smoothness and other properties of the model functions is encoded into the covariance function of the Gaussian process. Recently, Álvarez et al. (2009) introduced the idea of using Gaussian processes as non-parametric models for unknown input functions in physical models, which are formulated as differential equations (e.g. ODEs). As opposed to the classical models used in tracking, navigation and control applications, in this *latent force model* (LFM) approach, the input functions are not modeled as white noise processes, but instead their covariance structure is selected according to the machine learning approach, where the covariance structure is written in terms of unknown parameters that are estimated from data.

In this paper we show how non-linear latent force models can be represented as non-linear white noise driven state-space models, that is, partially observed non-linear stochastic differential equations (SDEs Jazwinski, 1970; Grewal & Andrews, 2001; Øksendal, 2003) and show how recently developed efficient non-linear Kalman filtering and smoothing based methods (Särkkä, 2007; Singer, 2008; Särkkä, 2010; Arasaratnam et al., 2010; Singer, 2011; Särkkä & Sarmavuori, 2012) can be used for inferring the state and parameters of these models. We compare the performance of the method to previously proposed Laplace-approximation and Markov chain Monte Carlo (MCMC) solutions (Lawrence et al., 2006; Titsias et al., 2009). The key advantages of the proposed approach over the previous ones is that i) its computational scaling is linear in the number of time steps and ii) it naturally handles stochasticity and non-linearity of the physical model.

Our main motivation for the proposed approach stems from an important real-world problem of long-term prediction of GPS satellite orbits (Seppänen et al., 2012). We show how the accuracy of orbit prediction can be improved by nonparametrically modelling forces unexplained by a deterministic non-linear physical orbit model.

## 2. Latent Force Models

*Latent force models* (LFMs) (Álvarez et al., 2009) are a relatively new modeling approach, combining mechanistic modeling principles (i.e., physical models) with non-parametric data-driven components. They have been successfully employed, for example, in ranked prediction of transcription factors (Honkela et al.). For instance, Lawrence et al. (2006) modelled the time-dependent expression levels $\{x_j(t)\}_{j=1}^N$ of $N$ genes with a system of first order ODEs

$$\frac{dx_j(t)}{dt} = B_j + \sum_{r=1}^{R} S_{j,r} g_j(u_r(t)) - D_j x_j(t), \quad j = 1, \ldots, N \quad (1)$$

where the driving processes $\{u_r(t)\}_{r=1}^R$ (representing the transcription factors, TFs) were given independent Gaussian process (GP) priors $u_r(t) \sim \mathcal{GP}(m(t), k_{u_r}(t, t'))$, $r = 1, \ldots, R$, where $m(t)$ and $k_{u_r}(t, t')$ were suitably chosen mean and covariance functions.

If the functions $g_j(\cdot)$ are linear, the model (1) is an instance of a *linear latent force model*. In such cases (see paper by Álvarez et al., 2009, for examples) the posterior inference on $x_j(t)$ and $u_r(t)$ is based on closed form computation of the covariance functions of $x_j(t)$, $dx_j(t)/dt$ and all the required cross covariances by solving the differential equation and then utilizing standard GP regression techniques.

However, in the case of non-linear $g_j(\cdot)$ (such as $g_j(u(t)) = e^{u(t)}$, ensuring the positivity of the forces effect) the ODE becomes non-linear. The standard GP techniques cannot anymore be applied since the needed covariance terms are analytically intractable. Inference in these models has been previously performed mainly by the Laplace method (Lawrence et al., 2006) and Markov chain Monte Carlo (MCMC) (Titsias et al., 2009). A severe limitation of these approaches is that they are based on the assumption that the likelihood of data can be written as an explicit function of the latent force process $\mathbf{u}(t)$ which can be evaluated (either approximately or exactly) in a computationally feasible manner.

## 3. SDE View of Latent Force Models

As discussed by Hartikainen & Särkkä (2010), GPs with certain stationary covariance functions (including the Matérn class) can be represented as solutions to linear time-invariant (LTI) SDEs (Øksendal, 2003). That is, we can formulate the GP priors on the $r = 1, \ldots, R$ components of $\mathbf{u}(t) = (u_1(t) \ldots u_R(t))^T$ as multivariate LTI SDEs of form

$$d\mathbf{z}_r(t) = \mathbf{F}_{z,r} \mathbf{z}_r(t) \, dt + \mathbf{L}_{z,r} \, d\beta_{z,r}(t) \quad (2)$$

where $\mathbf{z}_r(t) = (u_r(t) \; \frac{du_r(t)}{dt} \; \cdots \; \frac{d^{d_r-1} u_r(t)}{dt^{d_r-1}})^T$ and

$$\mathbf{F}_{z,r} = \begin{pmatrix} 0 & 1 & & \\ & \ddots & \ddots & \\ & & 0 & 1 \\ -a_r^0 & \cdots & -a_r^{p_r-2} & -a_r^{p_r-1} \end{pmatrix}, \mathbf{L}_{z,r} = \begin{pmatrix} 0 \\ \vdots \\ 0 \\ q_r \end{pmatrix}.$$

This dynamic model on $u_r(t)$ corresponds to a GP prior with a certain stationary covariance function when the coefficients $a_r^0, \ldots, a_r^{p_r-1}$, the diffusion constant $q_r$ and the dimensionality $p_r$ of $\mathbf{z}_r(t)$ are chosen appropriately.

Using this view on GP priors, the conversion of linear latent force models into linear-Gaussian state-space models was recently considered by Hartikainen & Särkkä (2011). In this paper, we consider conversion of non-linear latent force models into non-linear state-space models. Analogously to the linear case, a general non-linear latent force model can be formulated as a continuous-discrete system of form

$$d\mathbf{x}(t) = \mathbf{f}(\mathbf{x}(t), \mathbf{u}(t), t) \, dt,$$
$$d\mathbf{z}_r(t) = \mathbf{F}_{z,r} \mathbf{z}_r(t) \, dt + \mathbf{L}_{z,r} \, d\boldsymbol{\beta}_{z,r}(t), \; r = 1, \ldots, R$$

where $\mathbf{x}(t) \in \Re^M$ is the state ($M$ being the number of state components needed in representing the output processes $\{x_j(t)\}_{j=1}^N$ in a vector form), $\mathbf{u}(t) \in \Re^R$ the latent force processes and $\mathbf{f}(\cdot)$ the dynamic model function of output process $\mathbf{x}(t)$.

We can further simplify the notation by constructing an augmented system with state $\mathbf{x}_a(t)$ comprising of the output process and latent forces as $\mathbf{x}_a(t) = (\mathbf{x}(t), \mathbf{z}_1(t), \ldots, \mathbf{z}_R(t))^T$ with dynamics

$$d\mathbf{x}_a(t) = \mathbf{f}_a(\mathbf{x}_a(t), t) \, dt + \mathbf{L}_a(\mathbf{x}_a(t), t) \, d\boldsymbol{\beta}_a(t). \quad (3)$$

In the context of stochastic processes and filtering theory, the dynamic model function $\mathbf{f}_a$ is called the *drift function* and $\mathbf{L}_a$ the *dispersion matrix* weighting the $R$-dimensional Brownian motion $\boldsymbol{\beta}_a(t)$ with diffusion matrix $\mathbf{Q}$. Note that within this framework the model for $\mathbf{x}(t)$ does not need to be deterministic given $\mathbf{u}(t)$.

To complete the model specification we assume that observations at discrete time instants $t_1, \ldots, t_T$ can be modeled as

$$\mathbf{y}_k = \mathbf{h}_k(\mathbf{x}_a(t_k)) + \mathbf{r}_k, \quad k = 1, \ldots, T \quad (4)$$

where $\mathbf{h}(\cdot)$ is the measurement model function, $\mathbf{y}_k \in \Re^D$ is the measurement at time $t_k$ and $\mathbf{r}_k \sim \mathrm{N}(\mathbf{0}, \mathbf{R}_k)$ is the measurement noise.

## 4. Filtering and Smoothing of SDEs

Given continuous-discrete system of the form

$$\begin{aligned} d\mathbf{x}(t) &= \mathbf{f}(\mathbf{x}(t), t)\, dt + \mathbf{L}(\mathbf{x}(t), t) d\boldsymbol{\beta}(t) \\ \mathbf{y}_k &= \mathbf{h}_k(\mathbf{x}(t_k)) + \mathbf{r}_k, \quad k = 1, \ldots, T \end{aligned} \quad (5)$$

our aim is now to infer the *filtering* and *smoothing* distributions of the state $\mathbf{x}(t)$ at a time instant $t$:

$$p(\mathbf{x}(t) | \mathbf{y}_{1:k}), \quad t \in [t_k, t_{k+1}) \quad (6)$$

and

$$p(\mathbf{x}(t) | \mathbf{y}_{1:T}), \quad t \in [t_0, t_T], \quad (7)$$

where $\mathbf{y}_{1:i}$ is a shorthand notation for $\{\mathbf{y}_1, \ldots, \mathbf{y}_i\}$. In general, this is a difficult task that has been well studied in the field of stochastics and filtering theory (Jazwinski, 1970). The formal Bayesian continuous-discrete filter consists of separate prediction and update steps (Jazwinski, 1970), which are recursively iterated forward in time. On the prediction step we solve the probability density $p(\mathbf{x}(t_k) | \mathbf{y}_{1:k-1})$ of the state at time $t_k$, and on the update step, we use the Bayes' rule to update the density with latest observation as $p(\mathbf{x}(t_k) | \mathbf{y}_{1:k}) \propto p(\mathbf{y}_k | \mathbf{x}(t_k)) p(\mathbf{x}(t_k) | \mathbf{y}_{1:k-1})$. The prediction density $p(\mathbf{x}(t_k) | \mathbf{y}_{1:k-1})$ must be solved from the *Fokker–Planck–Kolmogorov (FPK) partial differential equation*, which is analytically intractable in general. Thus, to perform filtering and smoothing in practice approximate solutions must be sought. In this article, we consider Gaussian approximations for the filtering and smoothing distributions.

In the following we review the main steps of the Gaussian filtering and smoothing framework[1] which we consider appropriate for the type of models considered in this article. The general Gaussian filtering and smoothing framework is classical (see, e.g., Jazwinski, 1970; Maybeck, 1982), but here we utilize the recently developed sigma-point methods (Särkkä, 2007; Singer, 2008; Särkkä, 2010; Arasaratnam et al., 2010; Särkkä & Sarmavuori, 2012) for numerically solving the general continuous-discrete Gaussian filtering and smoothing equations. As shown by Singer (2011), the Gaussian filtering framework can also used for efficient evaluation of the likelihoods needed in parameter estimation methods.

[1] Matlab toolbox implementing the presented methods can be found from http://becs.aalto.fi/en/research/bayes/lfm/.

The filter works by recursively solving the following set of equations for time steps $k = 1, \ldots, T$:

- Solve mean and covariance of the predicted distribution $p(\mathbf{x}(t) | \mathbf{y}_{1:k-1}) \approx \mathrm{N}(\mathbf{m}(t_k^-), \mathbf{P}(t_k^-))$ by numerically integrating the differential equations[2]

$$\begin{aligned} \frac{d\mathbf{m}}{dt} &= \mathrm{E}[\mathbf{f}(\mathbf{x}, t)] \\ \frac{d\mathbf{P}}{dt} &= \mathrm{E}[(\mathbf{x} - \mathbf{m})\mathbf{f}^T(\mathbf{x}, t)] + \mathrm{E}[\mathbf{f}(\mathbf{x}, t)(\mathbf{x} - \mathbf{m})^T] \\ &\quad + \mathrm{E}[\mathbf{L}(\mathbf{x}(t), t)\, \mathbf{Q}\, \mathbf{L}(\mathbf{x}(t), t)^T], \end{aligned}$$

where the expectations are taken with respect to $\mathbf{x} \sim \mathrm{N}(\mathbf{m}(t), \mathbf{P}(t))$, and $t_k^-$ denotes the time instance "infinitesimally before the time $t_k$".

- Compute the approximate filtering distribution $p(\mathbf{x}(t_k) | \mathbf{y}_{1:k}) \approx \mathrm{N}(\mathbf{m}(t_k), \mathbf{P}(t_k))$ via the update (moment matching) equations

$$\begin{aligned} \mu_k &= \mathrm{E}[\mathbf{h}_k(\mathbf{x})], \\ \mathbf{S}_k &= \mathrm{E}[(\mathbf{h}_k(\mathbf{x}) - \mu_k)(\mathbf{h}_k(\mathbf{x}) - \mu_k)^T] \\ \mathbf{D}_k &= \mathrm{E}[(\mathbf{x} - \mathbf{m}_k^-)(\mathbf{h}_k(\mathbf{x}) - \mu_k)^T] \\ \mathbf{K}_k &= \mathbf{D}_k \mathbf{S}_k^{-1} \\ \mathbf{m}_k &= \mathbf{m}_k^- + \mathbf{K}_k\, (\mathbf{y}_k - \mu_k) \\ \mathbf{P}_k &= \mathbf{P}_k^- - \mathbf{K}_k \mathbf{S}_k \mathbf{K}_k^T. \end{aligned}$$

This is equivalent to the update step of a discrete-time filter, with definitions $\mathbf{m}_k^- \triangleq \mathbf{m}(t_k^-)$, $\mathbf{P}_k^- \triangleq \mathbf{P}(t_k^-)$ and $\mathbf{m}_k \triangleq \mathbf{m}(t_k)$, $\mathbf{P}_k \triangleq \mathbf{P}(t_k)$.

The approximate smoothing distributions $p(\mathbf{x}(t_k) | \mathbf{y}_{1:T}) \approx \mathrm{N}(\mathbf{x}(t_k) | \mathbf{m}^s(t_k), \mathbf{P}^s(t_k))$ can be obtained by recursively solving the following Kalman smoothing like equations for $k = T - 1, \ldots, 1$:

$$\begin{aligned} \mathbf{G}_{k+1} &= \mathbf{C}_k(t_{k+1})\, \mathbf{P}^{-1}(t_{k+1}^-) \\ \mathbf{m}^s(t_k) &= \mathbf{m}(t_k) + \mathbf{G}_{k+1}\, [\mathbf{m}^s(t_{k+1}) - \mathbf{m}(t_{k+1}^-)] \\ \mathbf{P}^s(t_k) &= \mathbf{P}(t_k) + \mathbf{G}_{k+1}\, [\mathbf{P}^s(t_{k+1}) - \mathbf{P}(t_{k+1}^-)]\, \mathbf{G}_{k+1}^T, \end{aligned}$$

where the cross covariance term $\mathbf{C}_k$ can be shown (Särkkä & Sarmavuori, 2012) to follow the differential equation

$$\frac{d\mathbf{C}_k}{dt} = \mathbf{C}_k\, \mathbf{P}^{-1}\, \mathrm{E}[\mathbf{f}(\mathbf{x}, t)\, (\mathbf{x} - \mathbf{m})^T]^T.$$

[2] These equations can be derived by applying the *Itô formula* to the FPK (see, e.g., Jazwinski, 1970).

This can be integrated alongside $\mathbf{m}(t)$ and $\mathbf{P}(t)$ during the filtering with negligible computation cost. An important thing to note here is that the time steps $t_k$ are not restricted to be the measurement time steps, that is, we can calculate the smoothed estimates at any time point inside the interval $t \in [t_0, t_T]$. On such steps the update equations of the filter are simply skipped.

The expectations in the equations above are taken over the approximating Gaussian distribution, and can be numerically computed with sigma-point and cubature integration methods (see, Särkkä, 2007; Singer, 2008; Särkkä, 2010; Arasaratnam et al., 2010; Särkkä & Sarmavuori, 2012). As was discussed by Singer (2011), the Gaussian filter also computes approximations to the conditional measurement likelihoods $p(\mathbf{y}_k|\mathbf{y}_{1:k-1}) \approx \mathrm{N}(\mathbf{y}_k|\mu_k, \mathbf{S}_k)$, which can be used in marginal likelihood based parameter learning via the factorization $p(\mathbf{y}_{1:T}|\theta) = \prod_{k=1}^{T} p(\mathbf{y}_k|\mathbf{y}_{1:k-1}, \theta)$, where $\theta$ denotes a vector of unknown parameters.

The weakness of the outlined inference scheme is that it assumes that the state distributions are approximately Gaussian. An alternative way of performing state inference would be to use a particle filter (Doucet et al., 2001), which approximates the posterior with sequential importance sampling. We could then use particle MCMC methods (Andrieu et al., 2010) for estimating the unknown parameters. The difficulty in using particle filters with SDEs is that when the state transition density cannot be evaluated in closed form, one is restricted to usage of the dynamic model as the importance distribution (Andrieu et al., 2010), which leads to inefficient sampling. Moreover, in the particle filter we need to solve the SDE numerically between the measurements for each sample separately, which is computationally very demanding.

## 5. Simulated Experiments

This section illustrates the performance of the proposed framework with two simulated examples.

### 5.1. Estimation of Transcription Factors

First we consider the TF model (1) with three different non-linear functions: (i) $g(u(t)) = \frac{e^{u(t)}}{\gamma + e^{u(t)}}$ (saturation), (ii) $g(u(t)) = \frac{1}{\gamma + e^{u(t)}}$ (repression) and (iii) $g(u(t)) = e^{u(t)}$ (exponential). The drift model in this case is linear and can thereby be solved approximately as a function of $u(t)$ (Lawrence et al., 2006), and thus it is possible to implement Laplace and MCMC for it.

For each non-linear function we generated 100 state trajectories for the time interval $t \in [0, 15]$ by the TF model (1) ($N = 3$, $R = 1$) with the latent force having a Matérn GP prior ($\nu = 3/2$). The model parameters were randomly generated for each state trajectory as $B_j \sim U(0, 0.1)$, $D_j \sim U(0, 2)$, $A_j \sim U(-0.1, 0.1)$, $S_j \sim U(0, 1)$, and were treated as fixed and known during the inference. For the saturation and repression functions we used the parameters $\gamma \in \{0.1, 0.5, 1\}$. For the GP prior we used the magnitude and length scale parameters $\sigma_{\mathrm{m}}^2 = 1$ and $l = 2$. Given the state trajectories we generated $T = 13$ equally placed observations for each of the $N$ outputs with additive Gaussian noise with variance $\sigma_{\mathrm{n}}^2 = 0.1^2$.

Our aim is to estimate the trajectory of the latent force given the generated observations. The estimation results are listed in Table 1 for Laplace approximation (LA), elliptical slice sampling (ESLS) (Murray et al., 2010) and a Gaussian continuous-discrete filter/smoother (GFS) with moment matching performed with the spherical cubature rule (Arasaratnam et al., 2010; Särkkä & Sarmavuori, 2012). With Laplace and ESLS we approximated the integral in the solution of the ODE by a Riemann sum with 363 grid points, and with GFS we used the 4th order Runge–Kutta method with 10 steps in integrating the moment equations. With ESLS we drew 100000 samples for $u(t)$, of which first 5000 were discarded as burn-in.

With Laplace approximation we used Newton's method with a simple step size adjustment procedure to find the mode. We excluded from the results all the cases where the Newton's method didn't stop after pre-specified number of iterations (300), as well as ones having RMSE (calculated over the 363 grid points) greater than $3\sigma_{\mathrm{m}}$ with any method, as this was taken to indicate that the method had diverged. The typical computation times for the data sets considered here were about 3 seconds with LA, 10 minutes with ESLS and 8 seconds with GFS.

From the results we can see that ESLS provides consistently good performance with all non-linear functions. GFS performs equally well with saturation and repression functions with all tested parameter configurations, but is slightly worse with the exponential function compared to Laplace and ESLS. However, Laplace had trouble in mode finding in several cases with the exponential function. With saturation and repression functions such problems were also present when $\gamma \in \{0.1, 0.5\}$, but with $\gamma = 1$ Laplace performed equally well with ESLS and GFS.

### 5.2. Tracking a Ballistic Target on Reentry

Next we consider tracking a ballistic target reentering the atmosphere with a sensor measuring the dis-tance to the target. In this example the drift and measurement models are non-linear, rendering Laplace and MCMC inapplicable for state inference. For simplicity we consider only a one dimensional case (that is, the target falls directly towards the ground), but the approach works similarly in a general 3D setting.

Table 1. **Results of the TF experiment.** The table lists the root mean square errors (RMSE) averaged over non-diverged simulations with all the tested non-linearities and methods, and the number of divergences (DIV). The results are listed in form RMSE (DIV).

|  | LA | ESLS | GFS |
|---|---|---|---|
| **Saturation** | | | |
| $\gamma = 0.1$ | 1.758 (0) | 0.737 (0) | 0.720 (0) |
| $\gamma = 0.5$ | 0.737 (0) | 0.484 (0) | 0.483 (0) |
| $\gamma = 1$ | 0.489 (0) | 0.483 (0) | 0.484 (0) |
| **Repression** | | | |
| $\gamma = 0.1$ | 1.933 (40) | 0.327 (0) | 0.374 (0) |
| $\gamma = 0.5$ | 1.267 (1) | 0.363 (0) | 0.367 (0) |
| $\gamma = 1$ | 0.483 (0) | 0.476 (0) | 0.474 (0) |
| **Exponential** | | | |
|  | 0.300 (55) | 0.290 (0) | 0.358 (9) |

The motion of the target is assumed to follow the equation (Li & Jilkov, 2001)

$$\begin{bmatrix} dr \\ dv \end{bmatrix} = \begin{bmatrix} -v \\ a(r,v,t) + g + u(t) \end{bmatrix} dt + \begin{bmatrix} q_r & 0 \\ 0 & q_v \end{bmatrix} \begin{bmatrix} d\beta_r \\ d\beta_v \end{bmatrix},$$

where $r$ is the altitude and $v$ the velocity of the target. We assume that the acceleration is caused by the drag force $a(r,v,t) = -\alpha \exp(-\gamma r) v^2$, where the exponentially decaying term models the air density, gravitational force $g = 9.8 \text{m/s}^2$ and an unknown force $u(t)$, which we assume to have a Matérn GP prior model ($\nu = 5/2, \sigma_m = 50\text{m/s}^2, l = 5\text{s}$). The drag force parameter was set $\alpha = 4.49 \times 10^{-4}$, and the air density scale to $\gamma = 1.49 \times 10^{-4}$. The noise parameters were set to $q_r = 50\text{m}/\sqrt{\text{s}}$ and $q_v = 10(\text{m/s})/\sqrt{\text{s}}$. The distance measurements were modeled as

$$y_k = \sqrt{s_x^2 + (s_y - r)^2} + r_k,$$

where $(s_x, s_y) = (30\text{km}, 30\text{m})$ is the position of the sensor, and measurement noise has variance $\sigma_n^2 = (30\text{m})^2$.

Starting from state $(r_0, v_0) = (65\text{km}, 3\text{km/s})$, we simulated the target trajectory on time interval $t \in [0, 30\text{s}]$ and generated 120 equally spaced measurements with the model above. Inference was performed with the Gaussian filtering and smoothing framework. We treated the drag force parameter $\alpha$ and the GP parameters $\sigma_m$ and $l$ as unknowns, which were determined by optimizing the marginal likelihood of data.

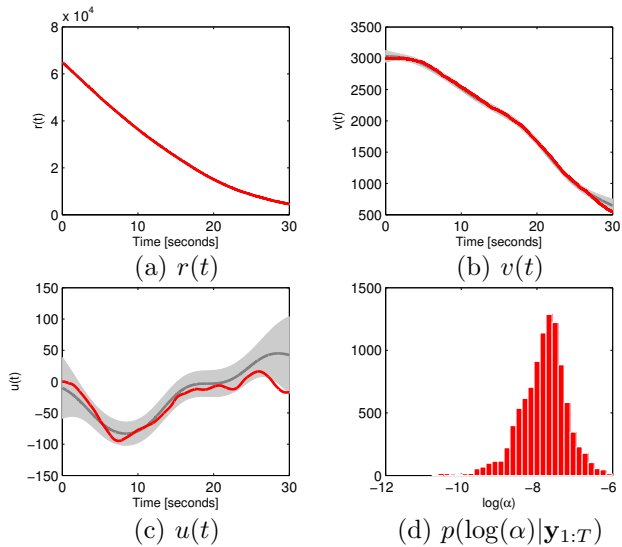

Figure 1. **Tracking a Ballistic Target on Reentry.** The panels (a), (b) and (c) show the smoothed estimates of $r(t)$, $v(t)$ and $u(t)$ (dark gray lines denoting the mean and light gray shade the 95% uncertainty) as well as the true simulated ones (red lines). Panel (d) shows a MCMC estimate of $p(\log(\alpha)|\mathbf{y}_{1:T})$.

Example results of estimating $r(t)$, $v(t)$ and $u(t)$ with optimized parameters are shown in panels (a), (b) and (c) of Figure 1. We also ran a MCMC inference for the unknown parameters using the negative log marginal likelihood provided by the filter as the energy function. Panel (d) shows the samples obtained for $\log(\alpha)$, peaking around the true value $(-7.7)$.

For comparison, we implemented also a particle filter for this model. We used a stochastic Runge–Kutta (strong order 1.5) to draw samples from the SDE. In practice, it required about 50000 particles to reach similar level of accuracy as with the Gaussing filtering scheme. In CPU time this took about 15 minutes, whereas Gaussian filtering takes only a few seconds.

## 6. GPS Satellite Orbit Prediction

As a real-world case study we consider the problem of predicting the orbit of a GPS satellite. Accurate modeling of the forces acting on a GPS satellite is needed in a number of applications and real-time applications require prediction of the orbit (Seppänen et al., 2012).

The equation of motion for the satellite can be written as a (vector) Markov model

$$\frac{d}{dt} \begin{bmatrix} \mathbf{r} \\ \mathbf{v} \end{bmatrix} = \begin{bmatrix} \mathbf{v} \\ \mathbf{a}(\mathbf{r}, t) + \mathbf{u}(\mathbf{r}, \mathbf{v}, t) \end{bmatrix},$$

where $\mathbf{a}(\mathbf{r}, t)$ is a deterministic model for the acceler-ation of the satellite, and $\mathbf{u}(\mathbf{r},\mathbf{v},t)$ represents acceleration terms caused by unknown forces acting on the satellite. Here, $\mathbf{r}$ and $\mathbf{v}$ represent the 3D position and velocity vectors of the satellite in an inertial coordinate system fixed to an arbitrary reference frame.

The deterministic acceleration of the motion model is

$$\mathbf{a}(\mathbf{r},t) = \mathbf{a}_\text{g} + \mathbf{a}_\text{moon} + \mathbf{a}_\text{sun} + \mathbf{a}_\text{srp}, \quad (8)$$

where $\mathbf{a}_\text{g}$, $\mathbf{a}_\text{moon}$, $\mathbf{a}_\text{sun}$ and $\mathbf{a}_\text{srp}$ are the accelerations due to Earth's gravitation, lunar gravitation, solar gravitation and solar radiation pressure, respectively.

When the asymmetrical mass distribution of the Earth is taken into account, its gravity potential $U$ can be written in the form of the spherical harmonics expansion (Montenbruck & Gill, 2005)

$$U(r, \lambda, \varphi) = \frac{\text{GM}_\text{E}}{r} \sum_{n=0}^{\infty} \sum_{m=0}^{n} \left[ \left(\frac{\text{R}_\text{E}}{r}\right)^n \text{P}_{nm}(\sin\varphi) \right.$$
$$\left. \left(C_{nm}\cos(m\lambda) + S_{nm}\sin(m\lambda)\right) \right]. \quad (9)$$

Here the potential $U$ is not only a function of satellite's radius $r$, but also the longitude $\lambda$ and latitude $\varphi$. The constant $\text{R}_\text{E}$ in this formula is the Earth's radius and the terms $\text{P}_{nm}$ are the associated Legendre polynomials of degree $n$ and order $m$. The coefficients $S_{nm}$ and $C_{nm}$ are experimentally determined constants, whose magnitude decreases very fast with increasing $n$ and $m$. Therefore, the potential can be approximated by taking into account only the first few terms. We used terms up to the degree and order 8. The acceleration due to Earth gravitation can be computed as gradient of the gravity potential $U$:

$$\mathbf{a}_\text{g} = \mathbf{R}^{-1} \nabla U, \quad (10)$$

where $\mathbf{R}$ is a suitable coordinate transformation matrix. For more details, see Montenbruck & Gill (2005) and Seppänen et al. (2012).

After Earth's gravitation the next biggest acceleration components in the satellite's equation of motion are caused by the gravitational forces of the Moon and the Sun. When dealing with Earth centered reference frame one has to compute the acceleration of the satellite in relation to the acceleration of the Earth. To compute this relative acceleration of the satellite caused by any celestial body, one can use the form

$$\mathbf{a}_\text{cb} = \text{GM}\left(\frac{\mathbf{r}_\text{cb} - \mathbf{r}}{\|\mathbf{r}_\text{cb} - \mathbf{r}\|^3} - \frac{\mathbf{r}_\text{cb}}{\|\mathbf{r}_\text{cb}\|^3}\right), \quad (11)$$

where M is the mass of the celestial body, $\mathbf{r}_\text{cb}$ is its position in the Earth centered inertial reference frame and $\mathbf{r}$ is the position of the satellite in the same reference frame. Applying this formula to Moon and Sun gives the accelarations $\mathbf{a}_\text{moon}$ and $\mathbf{a}_\text{sun}$ in Equation (8).

The last acceleration component in Equation (8), the solar radiation pressure, is a non-gravitational force whose exact form is not very well known. The main component of this force is pointing to the opposite direction from Sun. Furthermore, the amplitude of this force is almost constant, or the variations in the amplitude are rather small compared to its magnitude. Based on this information we can add a rough model for solar radiation pressure and later estimate the remaining parts of the force. The rough model is

$$\mathbf{a}_\text{srp} = -\alpha \frac{AU^2}{r_\text{sun}^2} \mathbf{e}_\text{sun}, \quad (12)$$

where $\mathbf{e}_\text{sun}$ is a unit vector from satellite to Sun, $AU$ is the astronomical unit and $r_\text{sun}$ is the distance from satellite to Sun. The satellite-specific constant amplitude $\alpha$ was batch estimated using half a year of position data of the satellite.

### 6.1. Modeling the Unknown Forces

As the first step in the modeling we would like to get a glimpse of what the unknown forces look like. To do this we assume separate GP smoothness priors for each component of $\mathbf{u}(\mathbf{r},\mathbf{v},t)$. Instead of placing the GP priors directly on the inertial coordinate system used in the integration, we place them on the radial, tangential and normal components of a RTN coordinate system with unit vectors

$$\mathbf{e}_R = \frac{\mathbf{r}}{\|\mathbf{r}\|}, \quad \mathbf{e}_T = \mathbf{e}_N \times \mathbf{e}_R, \quad \mathbf{e}_N = \frac{\mathbf{r} \times \mathbf{v}}{\|\mathbf{r} \times \mathbf{v}\|}.$$

Thus, the model for the unknown forces is

$$\mathbf{u}(\mathbf{r},\mathbf{v},t) = \mathbf{R}(\mathbf{r},\mathbf{v}) \begin{bmatrix} u_R(t) \\ u_T(t) \\ u_N(t) \end{bmatrix},$$

where $\mathbf{R}(\mathbf{r},\mathbf{v})$ is a matrix transforming RTN coordinates to the inertial coordinate system used in integration, and each of the latent forces $u_R, u_T$ and $u_N$ have Matérn GP priors. Overall, the model can be written in form (3), and thus be inferred with framework presented in Section 4.

Examples of smoothed force trajectories for ten days are shown as red lines in Figure 2. It is apparent that the force trajectories exhibit almost periodic, or *quasiperiodic* behavior, which can be utilized to improve predictions when modelled appropriately.

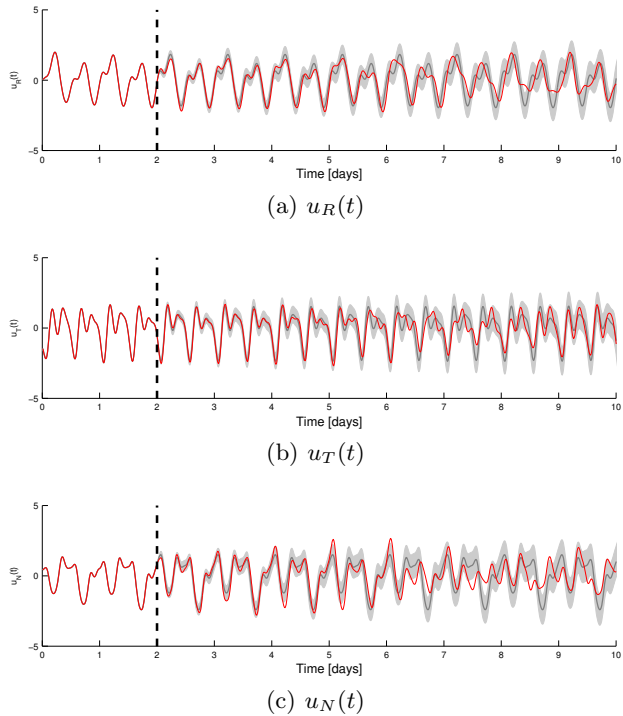

*Figure 2.* **GPS Satellite Prediction: estimated latent forces.** Red lines show the smoothed forces in RTN coordinates for 10 days with satellite 31. Dark gray lines denote the mean estimate of the constructed quasi-periodic force model and light gray the 95% uncertainty, when given the smoothed forces of first 2 days as observations, after which the model is used to predict the forces for the 8 rest days.

### 6.2. Quasi-Periodic Model for Latent Forces

To model the quasi-periodicities in the latent forces we use a stochastic resonator model, which previously has been used for modeling periodic phenomena in the brain (Särkkä et al., 2012). We model the periodic component as a superposition of resonators of the form

$$\frac{d^2 c_n(t)}{dt^2} = -(2\pi n f)^2 c_n(t) + w_n(t), \qquad (13)$$

where the additive white noise components $w_n(t)$ have spectral densities $q_n$. As shown by Särkkä et al. (2012), this model can be written in state space form

$$\begin{aligned} d\mathbf{x}(t) &= \mathbf{F}\,\mathbf{x}(t)\,dt + \mathbf{L}\,d\boldsymbol{\beta}(t) \\ u(t) &= \mathbf{H}\,\mathbf{x}(t) + b + \epsilon(t), \end{aligned} \qquad (14)$$

which is compatible with the framework presented in this article. In Equation (14) we have also included a bias term $b$ (which could also be time-varying) as the resonator model assumes the process to be zero-mean, and a white noise component $\epsilon(t)$ with spectral density $q_\epsilon$ to account for possible modeling errors.

### 6.3. Example Online Prediction Results

We now apply the constructed quasi-periodic latent force model to predict the satellite orbit, and compare the results to ones obtained with only the deterministic model. We consider an online prediction scenario, in which we observe the position and velocity of the satellite on certain time intervals, and between these intervals the models are used to provide predictions. We use 30 days of position data (collected every 15 minutes) of satellite 31 from the beginning of January 2010. The regions of observed data are illustrated with gray shades in Figure 3.

In the quasi-periodic latent force model we used 7 harmonic components to model $u_R(t)$, $u_T(t)$ and 10 in $u_N(t)$. As the period we used a little less than one day, which we observed to be a clear period in the estimated latent forces. The rest of the model parameters were optimized with respect to marginal likelihood, in which the smoothed mean estimate given by a Matérn GP model on a short time segment on the same satellite were treated as observed data. For inference in the actual GPS prediction with the latent force model we used the Gaussian continuous-discrete filter with the spherical cubature rule and moment integration by $4th$ order Runge–Kutta method with 80 steps between each observation. With the deterministic model the predictions were calculated by integrating the dynamics starting from the latest observation.

The errors in position estimates for both models are shown in Figure 3. It is evident that the modeling of periodicity reduces the position errors significantly. For example, in this particular case the position error of LFM after 15 days was less than 10% of that of the deterministic model. In fact, the amplitude of error even decreases during some time intervals, which might indicate the presence of some unexplained periodic forces acting on a longer time period. While we have here reported the predictions only with one satellite on a one-month time frame, the results are promising and certainly warrant further research.

### 7. Conclusion

In this article we have shown how non-linear latent force models can be represented as non-linear white noise driven state-space models. The resulting representation allows to apply efficient filtering and smoothing algorithms for state and parameter inference. The potential weakness of the approach is the underlying Gaussian approximation to the state posterior, but as we have shown there are many applications where the approach works well. The advantage of the approach

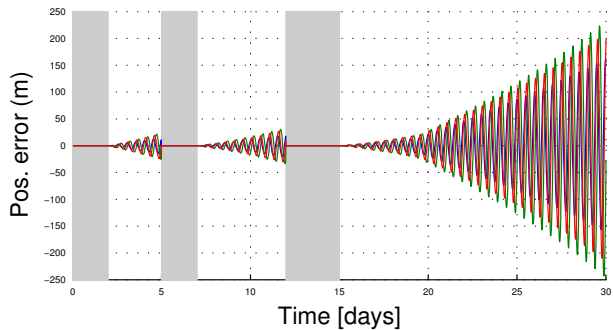

(a) Position error with deterministic model

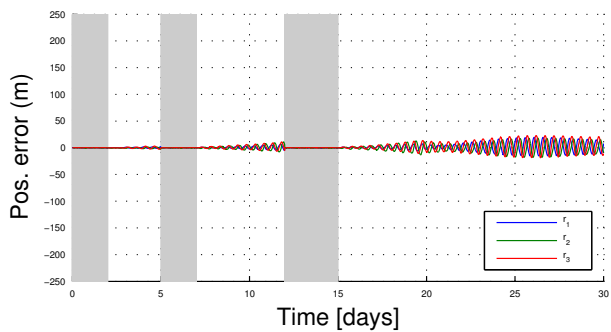

(b) Position error with quasi-periodic latent force model

Figure 3. **GPS Orbit Prediction: Errors in position estimates.** The Figure shows the errors in position estimates in the inertial coordinate system with (a) the deterministic model and (b) the latent force model given the data on shaded time intervals.

is that it is computationally light compared to, for example, pure particle filtering and MCMC based approximations, whose computational requirements can be infeasible for practical use.

ACKNOWLEDGEMENTS

The authors would like to thank Arno Solin, Janne Ojanen and anonymous reviewers for feedback that significantly improved the paper. The authors acknowledge financial support from Finnish Doctoral Programme in Computational Sciences, Finnish Foundation for Technology Promotion, Emil Aaltonen Foundation and Tampere Doctoral Programme in Information Science and Engineering.

## References

Álvarez, M. A., Luengo, D., and Lawrence, N. D. Latent Force Models. *JMLR: W&CP*, 5:9–16, 2009.

Andrieu, C., Doucet, A., and Holenstein, R. Particle Markov chain Monte Carlo methods. *JRSS B*, 72(3): 269–342, 2010.

Arasaratnam, I., Haykin, S., and Hurd, T. R. Cubature kalman filtering for continuous-discrete systems: Theory and simulations. *IEEE Transactions on Signal Processing*, 58(10):4977–4993, 2010.

Bar-Shalom, Y., Li, X-R., and Kirubarajan, T. *Estimation with Applications to Tracking and Navigation*. Wiley Interscience, 2001.

Doucet, A., de Freitas, N., and Gordon, N. (eds.). *Sequential Monte Carlo Methods in Practice*. Springer, 2001.

Grewal, M. S. and Andrews, A. P. *Kalman Filtering, Theory and Practice Using MATLAB*. Wiley Interscience, 2001.

Hartikainen, J. and Särkkä, S. Kalman Filtering and Smoothing Solutions to Temporal Gaussian Process Regression Models. In *Proc. MLSP*, pp. 379–384, 2010.

Hartikainen, J. and Särkkä, S. Sequential Inference for Latent Force Models. In *Proc. UAI*, 2011.

Honkela, A., Girardot, C., Gustafson, E. H., Liu, Y-H., Furlong, E.E.M., Lawrence, N. D., and Rattray, M. *PNAS*.

Jazwinski, A. H. *Stochastic Processes and Filtering Theory*. Academic Press, 1970.

Lawrence, N. D., Sanguinetti, G., and Rattray, M. Modelling transcriptional regulation using Gaussian processes. In *NIPS*, pp. 785–792, 2006.

Li, X. R. and Jilkov, V. P. Survey of maneuvering target tracking: II. Ballistic target models. In *Proc. SPIE*, volume 4473, 2001.

Maybeck, P. *Stochastic Models, Estimation and Control, Volume 2*. Academic Press, 1982.

Montenbruck, O. and Gill, E. *Satellite Orbits*. Springer, Berlin Heidelberg New York, 2005.

Murray, I., Adams, R. Prescott, and MacKay, D.J.C. Elliptical slice sampling. *JMLR: W&CP*, 9:541–548, 2010.

Øksendal, B. *Stochastic Differential Equations: An Introduction with Applications*. Springer, 6th edition, 2003.

Rasmussen, C. E. and Williams, C. K. I. *Gaussian Processes for Machine Learning*. MIT Press, 2006.

Särkkä, S. On unscented Kalman filtering for state estimation of continuous-time nonlinear systems. *IEEE Transactions on Automatic Control*, 52(9):1631–1641, 2007.

Särkkä, S. Continuous-time and continuous-discrete-time unscented Rauch-Tung-Striebel smoothers. *Signal Processing*, 90(1):225–235, 2010.

Särkkä, S. and Sarmavuori, J. Gaussian filtering and smoothing for continuous-discrete dynamic systems. 2012. Submitted.

Särkkä, S., Solin, A., Nummenmaa, A., Vehtari, A., Auranen, T., Vanni, S., and Lin, F.-H. Dynamic retrospective filtering of physiological noise in BOLD fMRI: DRIFTER. *NeuroImage*, 60(2):1517–1527, 2012.

Seppänen, M., Ala-Luhtala, J., Piché, R., and Martikainen, S. Autonomous prediction of GPS and GLONASS satellite orbits. *Navigation*, 2012. In press.

Singer, H. Nonlinear continuous time modeling approaches in panel research. *Statistica Neerlandica*, 62(1):29–57, 2008.

Singer, H. Continuous-discrete state-space modeling of panel data with nonlinear filter algorithms. *AStA*, 95 (4):375–413, 2011.

Titsias, M., Lawrence, N. D., and Rattray, M. Efficient sampling for Gaussian process inference using control variables. In *NIPS 21*, pp. 1681–1688. 2009.